\documentstyle[prl,floats,aps]{revtex}

\def\be{\begin{equation}}
\def\ee{\end{equation}}
\def\bea{\begin{eqnarray}}
\def\eea{\end{eqnarray}}

\begin{document}

\title{Numerical Evolution in time of curvature perturbations in Kerr black holes}
\author{Ram\'on L\'opez-Alem\'an}
\address{ Center for Gravitational Physics and Geometry,
Department of Physics,\\
The Pennsylvania State University, 104 Davey Lab, University Park, PA
16802.}
\date{November 2, 1999}
\maketitle
\begin{abstract}
In this paper I will review the basic features of the theory of curvature perturbations in Kerr spacetime, which is
customarily written in terms of gauge invariant components of the Weyl tensor which satisfy a perturbation equation
known as the Teukolsky equation. I will describe how to evolve generic perturbations about the Kerr metric and the
separable form of the wave solutions that one obtains, and the relation of the Teukolsky function to the energy of
gravitational waves emitted by the black hole. A discussion of a numerical scheme to evolve perturbations as a
function of time and some preliminary results of our research project implementing it for matter sources falling
into the black hole is included.
\end{abstract}

\section{Introduction}

The advent of the possibility of detecting the existence of gravitational waves with large interferometers (like the
LIGO or VIRGO projects now in their
 final construction stages) is a very exciting prospect for relativists and
astrophysicists. The weakness of the predicted signals requires the search for
 very strong sources. The best candidates to date for detectable sources are
black hole-black hole collisions. There is a very intense effort to numerically solve Einstein's equations in full
non-linear form to provide accurate templates to aid in the detection of the waveform from such a collision. Due to
the complexity of this task, this is a very ambitious problem still untractable even by state-of-the-art
supercomputers. Approximations like treating the last stages of the collision as a perturbation from a single
stationary black hole spacetime are very helpful both for aiding in the construction of a code for a full non-linear
evolution and to shed light on the physics of the collision itself.

In black hole perturbation theory we want to linearize the Einstein equations for gravitational (and other matter
fields) about one of the known stationary solutions. This is a simplification over solving the full non-linear set
of equations, although the resulting equations can be quite messy in algebraic terms. But in this way we can evolve
small departures from the isolated black hole spacetime due to the presence of outside sources like particles,
gravitational waves or even collisions with other black holes (in the ``close limit'' where the two objects can be
regarded as one distorted black hole).

\pagestyle{myheadings}
\markright{R. L\'OPEZ-ALEM\'AN}

A straightforward way to do this is to consider metric perturbations. This is analogous to what is done in the
linearized treatment of gravity when the metric is written as $g_{ab}=\eta_{ab} + h_{ab}$, where $h_{ab}$ is small
compared to the Minkowski metric. In the late 50's, Regge and Wheeler\cite{R-W} were the first to do this by
linearizing the vacuum Einstein equations about the Schwarzschild metric and using the symmetries inherent in the
background metric to separate the angular and radial parts of the equation. They expand the solution in tensor
spherical harmonics, leaving a rather simple Schr\"{o}dinger-like equation to solve for the radial part of the
odd-parity set of perturbations.

The Regge-Wheeler approach, even when it has been quite useful for the Schwarzschild case, involves quite a lot of
complex algebra, particularly when treating even-parity perturbations. It is unfeasible to extend it to the more
complex Kerr geometry. In addition, it suffers some of the same drawbacks as the linearized theory, since the
perturbations are coordinate gauge dependent. If one does an infinitesimal coordinate translation $x_a \rightarrow
x_a + \epsilon \xi_a$ , the metric function $h_{ab}$ transforms as

\begin{equation}
  h_{ab} \rightarrow h_{ab} - 2 \nabla_{(b} ~\xi_{a)}
\end{equation}

The use of gauge dependent perturbations is warranted if one asks the right questions about global properties of the
gravitational radiation and avoids trying to say anything about properties of the gravitational field at local
events in the spacetime. Still one would be able to avoid all ambiguities if only gauge independent perturbations
are considered.

\subsection{Special algebraic character of the Kerr metric}

The Kerr metric\cite{Kerr} is the axisymmetric solution to Einstein's field equations that describes the spacetime
outside a stationary, rotating black hole. It is a
 type II-II in the Petrov classification scheme, which means that out of the
four principal null directions that all stationary spacetimes can have it has two distinct principal null
directions, (since the other two coincide with these).

These directions are null vectors that satisfy the following condition :

\begin{equation}
  k^b k^c C_{abc[d} ~k_{e]}= 0
\end{equation}

\noindent where the $C_{abcd}$ is the Weyl tensor.

This fact will be very convenient when one treats curvature perturbations using the Newman-Penrose
formalism\cite{NP}, since if one selects these principal null directions as the basis for the NP tetrad, one gets
enough relations between the various spin coefficients to make the resulting system of equations easily solvable.

\subsection{The Newman- Penrose formalism}

The Newman-Penrose formalism was developed to introduce spinor calculus into general relativity. It is a special
instance of tetrad calculus\cite{Wald}. Let us present a brief summary of the basic ideas behind the use of the
formalism which will be used in deriving the Teukolsky equation.

One starts by introducing a complex null tetrad \{  {\boldmath$l,n,m,m^*$} \} at each point in spacetime which
consists of two real null vectors {\boldmath $l,n$} and one complex spacelike vector {\boldmath $m$}. These should
satisfy the orthonormality relations

\boldmath
\begin{equation}\label{eq:onr}
  l \cdot n = m \cdot m^* = 1
\end{equation}
\unboldmath

\noindent with all other products being zero. Then $g_{ab} = -2 [l_{(a} n_{b)} - m_{(a} m^*_{b)}]$. Now one defines
four directional derivative operators along the tetrad directions

\bea D = l^a \nabla_a, &  & \Delta = n^a \nabla_a, \nonumber \\ \delta = m^a \nabla_a, &  & \delta^* = m^{*a}
\nabla_a. \label{eq:dvops} \eea

The basic quantities of the formalism are the {\it spin coefficients}, of which there are twelve complex ones :

\bea &{}\alpha = \frac {1}{2}~ (n^a m^{*b} \nabla_b l_a - m^{*a} m^{*b} \nabla_b m_a ), \nonumber \\ &{}\beta =
\frac {1}{2}~ (n^a m^{b} \nabla_b l_a - m^{*a} m^{b} \nabla_b m_a ), \nonumber \\ &{}\gamma = \frac {1}{2}~ (n^a
n^{b} \nabla_b l_a - m^{*a} n^{b} \nabla_b m_a ), \nonumber \\ &{}\epsilon = \frac {1}{2}~ (n^a l^{b} \nabla_b l_a -
m^{a} l^{b} \nabla_b m_a ), \nonumber \\ &{}\lambda = - m^{*a} m^{*b} \nabla_b n_a ,~~ \mu = - m^{*a} m^b \nabla_b
n_a, \nonumber \\ &{}\nu = - m^{*a} n^b \nabla_b n_a ,~~ \pi = - m^{*a} l^b \nabla_b n_a, \nonumber \\ &{}\kappa =
m^a l^b \nabla_b l_a ,~~ \rho = m^a m^{*b} \nabla_b l_a, \nonumber \\ &{}\sigma = m^a m^b \nabla_b l_a ,~~ \tau =
m^a n^b \nabla_b l_a \eea

The whole set of field equations in the formalism come by writing the Ricci and Bianchi identities using these
coefficients, and they take the place of the Einstein equations.

All ten independent components of the Weyl tensor can be written as five complex scalars:

\begin{eqnarray}
 &\psi_0 = - C_{abcd}~ l^a m^b l^c m^d \nonumber \\
 &\psi_1 = - C_{abcd}~ l^a n^b l^c m^d \nonumber \\
 &\psi_2 = - \frac{\,1}{\,2}~ C_{abcd} (l^a m^b l^c m^d + l^a n^b m^c m^{*d})
 \nonumber \\
 &\psi_3 = - C_{abcd}~ l^a n^b m^{*c} n^d \nonumber \\
 &\psi_4 = - C_{abcd}~ n^a m^{*b} n^c m^{*d}
\eea

To do perturbation calculations one specifies the perturbed geometry by introducing slight changes in the tetrad
like {\boldmath $l = l^A + l^B , n = n^A + n^B$}, etc. Here the {\it A} terms are the unperturbed values and the
{\it B} ones the small perturbation. Then, all the Newman-Penrose spin coefficients and other quantities can be also
written in a similar fashion
: $\psi_4 = \psi _{4}^{A} + \psi _{4}^{B}$, etc. The perturbation equations come
from the Newman-Penrose set by keeping {\it B} terms only up to first order.

\section{The Teukolsky equation}

In 1973, Teukolsky\cite{teuk} used the Newman-Penrose formalism to the special case of the background geometry of a
II-II type, (the Kerr or Schwarzschild black holes are of this type). In this way he was able to deduce the
linearized equations for full dynamical perturbations of the hole that could handle changes in its mass and angular
momentum, interaction with accreting test matter or distant massive objects, etc.

This approach has many important advantages. First, it turns out rather surprisingly that the equations are
separable, so that by Fourier transforming and expressing the solution as a series expansion of {\it spheroidal
harmonics} one ends up with having to solve just an ordinary differential equation for the radial part just like in
the Regge-Wheeler case (in fact, the solutions are related to each other by a transformation operator, as we will
discuss later). Second, since for gravitational perturbations the dependent variable will be constructed out of the
Weyl tetrad components $\psi_0$ and $\psi_4$, this will describe {\it gauge independent} perturbations, because
these are gauge-invariant quantities ( for more details on how to determine the gauge dependence of perturbations in
general, one can look up the review by Breuer\cite{Bre} ).

When one chooses the \boldmath $l$ and $n$ \unboldmath vectors of the unperturbed tetrad along the repeated
principal null directions of the Weyl tensor, then

\bea & \psi^A_0 = \psi^A_1 = \psi^A_3 = \psi^A_4 = 0 \nonumber \\ & \kappa^A = \sigma^A = \nu^A = \lambda^A = 0 \eea

By collecting the Newman-Penrose equations that relate $\psi_0, \psi_1$ and $\psi_2$
 with the spin coefficients and tetrad components of the stress-energy tensor, and linearizing
 about the perturbed values, Teukolsky gets (after some algebra) the following decoupled
 equation for the perturbed $\psi_0$ :

\bea
  &&{}[(D - 3 \epsilon + \epsilon^* - 4 \rho - \rho^*)(\Delta - 4 \gamma + \mu) \nonumber \\
  &&{} -(\delta
  + \pi^* - \alpha^* - 3 \beta - 4 \tau) (\delta^* + \pi - 4 \alpha) - 3 \psi_2]
  \psi_0^B = 4 \pi T_0
\eea where \bea &&{}T_0 = (\delta + \pi^* -\alpha^*-3 \beta-4 \tau)[(D-2 \epsilon-2 \rho^*)T^B_{lm} \nonumber \\
&&{}- (\delta + \pi^*-2 \alpha^* - 2\beta)T^B_{ll}]+(D-3 \epsilon + \epsilon^*-4 \rho - \rho^*)\nonumber \\
&&{}\times [(\delta+2 \pi^* - 2 \beta)T^B_{lm} - (D+2 \epsilon +2 \epsilon^*- \rho^*)T^B_{mm}]~ \eea

Since the full set of NP equations remains invariant under the interchange \boldmath $l \leftrightarrow n$, $m
\leftrightarrow m^*$ \unboldmath (this is the basis of the GHP method \cite{GHP}), then by applying this
transformation one can derive a similar equation for $\psi^B_4$ :

\bea &&{}[(\Delta + 3 \gamma - \gamma^* + 4 \mu + \mu^*)(D + 4 \epsilon - \rho) \nonumber
\\&&{}  - (\delta^* - \tau^* + \beta^* +3 \alpha + 4 \pi)(\delta - \tau + 4 \beta) - 3
\psi_2]\psi^B_4=4 \pi T_4 \eea

where

\bea &&{}T_4 = (\Delta + 3 \gamma - \gamma^* + 4 \mu + \mu^*)[(\delta^* - 2 \tau^* + 2 \alpha)T^B_{nm^*} \nonumber
\\ &&{}- (\Delta + 2 \gamma - 2 \gamma^* + \mu^*)T^B_{m^*m^*}] + (\delta^* - \tau^* + \beta^* + 3 \alpha + 4 \pi)
\nonumber \\ && \times [(\Delta + 2 \gamma + 2 \mu^*)T^B_{nm^*} - (\delta^* - \tau^* + 2 \beta^* + 2 \alpha)
T^B_{nn}] \label{T4} \eea

In a similar way, one can define tetrad components of the electromagnetic field tensor

\bea \Phi_0 = F_{\mu \nu} l ^{\mu} m^{\nu},& \Phi_1 = \frac{\,1}{\,2}~ F_{\mu \nu}(l^{\mu} n^{\nu} + m^{* \mu}
m^{\nu}),& \Phi_2 = F_{\mu \nu}m^{* \mu} n^{\nu} \eea

and get similar decoupled equations for $\Phi_0$ and $\Phi_2$. One can try these ideas with neutrino and scalar
fields also. So if one now writes the tetrads in Boyer-Lindquist \cite{BL} coordinates ${t,r,\theta,\phi}$ (after
using the gauge freedom to set up the spin coefficient $\epsilon = 0$) \cite{teuk} they become

\bea & l^{\mu} = [(r^2 + a^2)/\Delta, 1, 0, a/\Delta],~~ n^{\mu} = [r^2 + a^2, -\Delta, 0, a]/(2 \Sigma), \nonumber
\\ & m^{\mu} = [i a sin \theta, 0, 1, i/sin \theta]/(\sqrt{2} (r + i a cos \theta)) \eea

where {\it aM} is the angular momentum of the black hole, $\Sigma = r^2+ a^2 cos^2 \theta$, and $\Delta = r^2 - 2 M
r +a ^2$ (note that before one of the differential operators $n^{\mu} \partial_{\mu}$ was given the symbol $\Delta$,
but from now on it will have the conventional sense described here). With these expressions one can now write
explicitly the spin coefficients and $\psi_2$. Then it turns out that one can write all the decoupled equations for
test scalar fields ($s = 0$), a test neutrino field ($s = \pm \frac{\,1}{\,2}$), a test electromagnetic field ($s =
\pm 1$) or a gravitational perturbation ($s = \pm 2$) as a single master equation which is the famed{\it Teukolsky
equation} :

\bea &&{}\left [ \frac{(r^2+a^2)^2}{\Delta} - a^2 sin^2 \theta \right ] \frac{\partial^2 \psi}{\partial t^2} +
\frac{4 M a r}{\Delta} \frac{\partial^2 \psi}{\partial t
\partial \phi} + \left [ \frac{a^2}{\Delta} - \frac{1}{sin^2 \theta} \right ] \frac{\partial^2
\psi}{\partial \phi^2} \nonumber \\&& - \Delta^{-s} \frac{\partial}{\partial r} \left ( \Delta^{s+1}\frac{\partial
\psi}{\partial r}\right ) - \frac{1}{sin \theta} \frac{\partial}{\partial \theta} \, \left ( sin \theta
\frac{\partial \psi}{\partial \theta} \right ) -2 s \left [\frac{a (r-M)}{\Delta} + \frac{i \, cos \theta}{sin^2
\theta} \right ] \frac{\partial \psi}{\partial \theta} \nonumber \\ && - 2 s \left [ \frac{M(a^2 - r^2)}{\Delta} - r
- i\, a\, cos \theta \right ] \frac{\partial \psi}{\partial t} + \left [ s^2 cot^2 \theta - s \right ] \psi = 4 \pi
\Sigma T \eea

For the case that interest us, which is where the perturbations are to be interpreted as gravitational radiation
that can be measured at infinity, the value for s = -2, $\psi = \rho^{-4} \psi_4^B$, where $\rho = -1/(r - i \,a \,
cos \theta)$ in the coordinates we are using, and $T = 2 \rho^{-4} T_4$.

As mentioned before, this equation turns out to be separable. If one writes the Teukolsky perturbative function as
$\psi = e^{-i\, \omega t} e ^ {i \, m \phi} S(\theta) R(r)$ then the Teukolsky equation separates into a radial part
and an angular part. The angular equation for $S(\theta)$ has as a complete set of eigenfunctions the ``spin
weighted spheroidal harmonics'' \cite{Stew} of weight $s$ . The radial part can be written as

\bea \left ( \frac{d}{dr} p \frac{d}{dr} + p^2 U\right ) R = p^2 T \eea

In here, $p(r) = (r^2 - 2 M r)^{-1}$ and the effective potential is given by $U =~(1 - \frac{2 M}{r})^{-1}[(\omega
r)^2 - 4 i \omega (r - 3 M)] - (l-1)(l+2)$, while T is the source term previously defined for $s=-2$. (This is for
the simplified $a = 0$ case. A slightly more complicated version depending on the value of {\it a} for the fully
rotating case can be found as equation (4.9) of Teukolsky\cite{teuk}.)

When the angular momentum parameter $a$ goes to zero, the Kerr metric goes to the Schwarzschild one and the
Teukolsky equation becomes the Bardeen-Press equation\cite {BP}.

There is of course considerable interest in computing the energy carried off by outgoing waves at infinity due to
the evolution of the perturbations from some initial data. The non-trivial information about outgoing waves at
infinity is carried by the $\psi_4^B$ tetrad component. In principle, it is possible to use the solution for
$\psi_4^B$ to solve the complete Newman-Penrose set of equations for the perturbations in the metric. So, for
outgoing waves with frequency $\omega $

\bea \psi_4^B = -\omega^2 (h_{\theta \theta}^B - i \, h_{\theta \phi}^B)/2 \eea

Therefore,

\begin{equation}
\frac{d^2 E^{(out)}}{dt d\Omega} = \; \stackrel {\displaystyle \lim}{\scriptscriptstyle r \rightarrow \infty}
\frac{r^2 \omega^2}{16 \pi}\left [ (h_{\theta \theta}^B)^2 +(h_{\theta \phi}^B)^2 \right ]= \;\stackrel
{\displaystyle \lim}{\scriptscriptstyle r \rightarrow \infty} \frac{r^2}{4 \pi \omega^2} |\, \psi_4^B \, |^2
\end{equation}

\section{Calculation of perturbations from infalling particles}

Both the Regge-Wheeler and the Teukolsky equation have been used extensively in one of the simple test cases for
this perturbation formalism : that of a particle of mass $\mu \ll M$ falling into a stationary, isolated black hole.
That leads later on to extensions like considering the deformation and internal dynamics of an infalling star,
accretion disks \cite{Papadou} , and hopefully the late stages of a black hole collision in a not-too-distant
future.

The first such calculation using a Green's function technique to integrate the recently derived Zerilli equation
(the even-parity counterpart of the Regge-Wheeler equation) was done by Davis, Ruffini, Press and Price \cite{DRPP}.
They computed for the first time the amount of energy that was given out as gravitational waves by a particle
falling radially from infinity into a Schwarzschild black hole, and they found that it radiated $\Delta E = 0.0104
\, \mu^2 / M$ in geometrized units. The radiation from the {\it l\/} =2 multipole dominates the spectrum and is
peaked at $\omega $~= 0.32 $M^{-1}$ , which just a little below the fundamental resonant frequency for the black
hole.

Sometime later, Ruffini \cite{Ruffini} treated the case of a particle falling radially from infinity but with
non-zero initial velocity, with the result that the increase in radiated energy was minimal. More general treatments
were attempted by Detweiler \& Szedenits \cite{DS} which examined infall trajectories with nonzero angular momentum.
Considerable increases in the emitted gravitational radiation are seen as the normalized angular momentum of the
trajectory $J / \mu M$ increases from 0 to close to $4 M$ (where the particle approaches a marginally bound,
circular orbit). Increases in $\Delta E$ by a factor of 50 are found at the high {\it J }end.

The first calculation involving a Kerr black hole was carried out by Sasaki and Nakamura \cite{SaNam} who considered
a particle falling radially along the symmetry axis of the hole. Several studies (all in the frequency domain) have
been carried out dealing with infall in the equatorial plane (and the effect of the rotational frame dragging)
\cite{Koj1}, infall trajectories with finite angular momentum \cite{Koj2}, and quite a few detailed simulations of
the gravitational waves emitted by a particle in a bound orbit around the black hole. \cite{PaS}

\section{Conclusions and Current Research}

The Teukolsky equation is a powerful and very convenient way to deal with gauge invariant curvature perturbations of
both the Kerr and Schwarzschild metrics. It has a very nice, separable mathematical structure and is amenable to
robust numerical integration. Many interesting results have been derived leading up to the ideal situation of using
it to treat the close limit of a generic rotating black hole collision and the evolution of gravitational waveforms
from it.

Up to now, basically all treatments have been based on the separability of the equation and calculate the energy and
waveforms for the first few {\it l} multipoles of the spheroidal harmonics expansion once the radial part has been
dealt with. However, for the purpose of detecting the gravitational waves from the inspiral collision of a binary
black hole system using laser interferometers one would like to obtain the time integration of the full Teukolsky
equation once we have started from reasonable initial data describing the two holes in close proximity to each
other.

Krivan {\it et al.} \cite{Laguna} have devised a procedure to evolve perturbations in time from generic initial data
using the Teukolsky equation. Their method analyzes the radiation at infinity by dealing with the $s = -2$ version
of the equation. They avoid fully separating the Teukolsky function and use the ansatz

\begin{equation}\label{ansatz}
  \psi \equiv e^{i\, m \phi} r^3 \Phi (t,r^*, \theta)
\end{equation}

Then the equation is rewritten as a first order matrix equation and numerically integrated. It has given encouraging
results in treating scalar fields, scattering of gravitational waves and analysis of quasi-normal ringing and power
law tails of the outgoing radiation. This demonstrated the feasibility of this numerical approach for the
homogeneous Teukolsky equation.

We are now working on the case of the infalling particle to again try to reproduce the Davis {\it et al.} results.
The idea is to simulate the stress-energy tensor for a point particle

\begin{equation}
  T^{\alpha \beta}(x) = \int \mu \,u^{\alpha} u^{\beta} \delta^4(x - z(\tau)) d\tau
\end{equation}

\noindent by using very narrow gaussian distributions in place of the spatial Dirac delta function remaining after
the integration. In here, $z(\tau)$ is the geodesic trajectory of the falling particle of mass $\mu$, and
$u^{\alpha}$ is its four-velocity. Then one contracts with the null tetrad vector and inputs this into equation
(\ref{T4}). Since this is a 2-d code that evolves the Teukolsky equation in an $(r, \theta)$ grid, one must
carefully keep track of the $\phi$ dependence and all its derivatives implicit in equation (\ref{T4}) and then
Fourier expand this piece of the source term to match the $\phi$ dependence of the expression for $\psi$ in equation
(\ref{ansatz}).

We have had some encouraging preliminary results with the outlined procedure described here for the simplified case
of a particle falling along a radial trajectory to a non-rotating black hole, and we are actively working to modify
the code to tackle more general and interesting problems in perturbations of rotating black holes in the presence of
infalling matter.

\section*{Acknowledgements}

I wish to thank my advisor J. Pullin, and P. Laguna of the PSU Dept. of Astronomy and Astrophysics for their helpful
comments and suggestions. This work was supported in part by the Penn State Graduate School Academic Computing
Fellowship Program, and by the National Science Foundation via grants NSF-PHY-9423950 and NSF-PHY-9800973.

\end{document}